**Research Article**

Mohammadreza Saemian[+], Livia Del Balzo[+], Djamal Gacemi, Yanko Todorov, Etienne Rodriguez, Olivier Lopez, Benoit Darquié, Lianhe Li, Alexander Giles Davies, Edmund Linfield, Angela Vasanelli[*], and Carlo Sirtori

[+] These authors contributed equally to this work

# Ultra-sensitive heterodyne detection at room temperature in the atmospheric windows

**Abstract:** We report room temperature heterodyne detection of a quantum cascade laser beaten with a local oscillator on a unipolar quantum photodetector in two different atmospheric windows, at 4.8 μm and 9 μm. A noise equivalent power of few pW is measured by employing an active stabilization technique in which the local oscillator and the signal are locked in phase. The measured heterodyne noise equivalent power is six orders of magnitude lower than that obtained with direct detection.

## 1 Introduction

Promoting uncooled sensitive and high bandwidth photodetection in the mid (3μm < λ< 8μm) and long wave (8μm < λ< 15μm) infrared range is a challenging problem essential for several applications, spanning from high-bit-rate free-space communications in the atmospheric windows [1] to quantum metrology [2,3]. In addition, a great research effort has been dedicated into LiDAR systems, in particular for automotive applications [4] and for the simultaneous measurement of velocity and distance of molecules in the atmospheric windows [5]. For the development of LiDAR technology in the wavelength ranges considered here, the detection of very weak signals is a major challenge. Heterodyne detection, where a weak signal is coherently mixed with a powerful local oscillator on a fast detector, is a promising technique to address this issue. We propose and demonstrate the implementation of this technique in the thermal infrared range by using unipolar quantum optoelectronic devices, such as quantum cascade lasers (QCLs) [6], quantum well infrared photodetectors (QWIPs) [7] and quantum cascade detectors (QCDs) [8]. Indeed, QCLs feature today continuous-wave single-mode emission with output powers over 100 mW and are, therefore, powerful sources of coherent radiation that can be used as local oscillators [9-12]. On the other hand, QWIPs and QCDs show a high optical saturation power [13] and large frequency bandwidth, exceeding 100 GHz [14], making them ideal heterodyne receivers. The performances of such detectors in terms of signal-to-noise ratio, frequency bandwidth and temperature operation have been recently improved by inserting them into metamaterial architectures that insure a good coupling with free space radiation and reduce the electrical area [15-17]. Additionally, unipolar quantum optoelectronic devices profit of the very mature semiconductor packaging technology that makes them perfect chips for compact instruments and optical systems. The sensitivity of the heterodyne detection depends on the stability of the beat-note between the signal and the local oscillator, because the signal to noise ratio is directly proportional to the integration time. When free-running QCLs are used in heterodyne systems, the noise equivalent power (NEP) is set by the different sources of noise limiting their frequency stability. Typical free running linewidths are of few MHz due to the noise arising from electrical current [18,19] and other low frequency fluctuations that set the value of the noise equivalent power (NEP). In this work, we mitigate the effect of these fluctuations thanks to an active stabilization technique in which the local oscillator and the signal are locked in frequency and phase by the injection of a correction current derived from a phase-lock-loop (PLL) [2,20]. As a result, we measure a noise equivalent power as low as few pW, more than one order of magnitude lower than previous demonstrations [16]. The paper is organized as follows. After presenting our experimental set-up in section 2, we discuss in section 3 the sensitivity of optical heterodyne detection with free-running QCLs by measuring the associated NEP. Then, in section 4, we focus on the effect of the stabilization technique on the reduction of the NEP. Finally, in section 5, we discuss possible ways to further improve our system.

[*]**Corresponding author: Angela Vasanelli**, Laboratoire de Physique de l'École Normale Supérieure, ENS, Université PSL, CNRS, Sorbonne Université, Université Paris Cité, 75005 Paris, France; angela.vasanelli@ens.fr (https://orcid.org/0000-0003-1945-2261)
**Mohammadreza Saemian, Livia Del Balzo, Djamal Gacemi, Yanko Todorov, Etienne Rodriguez, Carlo Sirtori:** Laboratoire de Physique de l'École Normale Supérieure , ENS, Université PSL , CNRS,Sorbonne Université, Université Paris Cité, 75005 Paris, France; mohammadreza.saemian@phys.ens.fr (https://orcid.org/0009-0009-7740-6977), livia.delbalzo@phys.ens.fr (https://orcid.org/0009-0007-3262-6330), djamal.gacemi@phys.ens.fr (https://orcid.org/0000-0003-0334-1815), yanko.todorov@phys.ens.fr ( https://orcid.org/0000-0002-2359-1611), etienne.rodriguez@gmail.com (https://orcid.org/0000-0002-2397-1956), carlo.sirtori@ens.fr; https://orcid.org/0000-0003-1817-4554
**Olivier Lopez, Benoit Darquié:** Laboratoire de Physique des Lasers, CNRS, Université Sorbonne Paris Nord, 93430 Villetaneuse, France; olivier.lopez@univ-paris13.fr, benoit.darquie@univ-paris13.fr, ,
**Lianhe Li, Alexander Giles Davies, Edmund Linfield:** School of Electronics and Electrical Engineering, University of Leeds, Woodhouse Lane, Leeds LS2 9JT, UK; L.H.Li@leeds.ac.uk ( https://orcid.org/0000-0003-4998-7259) , G.Davies@leeds.ac.uk (https://orcid.org/0000-0002-1987-4846) , G.Davies@leeds.ac.uk, e.h.linfield@leeds.ac.uk (https://orcid.org/0000-0001-6912-0535 )

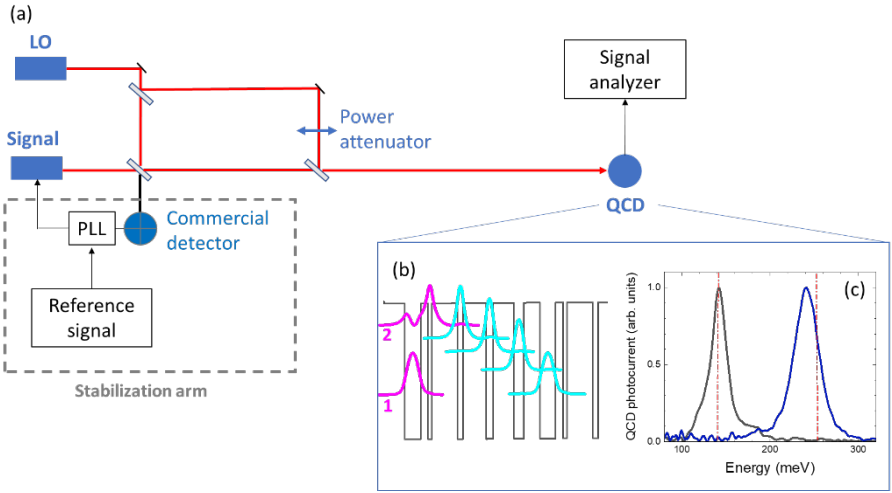

**Figure 1.** (a) Diagram of the experimental setup. (b) Conduction band profile and square moduli of the relevant wavefunctions of one period of the 4.8 μm QCD used in the experiment. (c) Room temperature photocurrent spectra of the two QCDs. The dashed red lines indicate the laser emission energies. LO: local oscillator; PLL: phase-lock loop; QCD: quantum cascade detector.

## 2 Experimental setup

The experimental setup for optical heterodyne detection is presented in Figure 1(a). The heterodyne beatnote is produced by two QCLs, the first is the local oscillator (LO), the second is referred as signal. Two different set-ups have been realized, operating in two different atmospheric windows, at 4.8 μm and 9 μm. Each set-up employs a couple of commercial distributed feedback (DFB) QCLs operating in continuous wave at room temperature (from Ad Tech Optics at λ=4.8 μm and from Thorlabs at λ=9 μm). Two different optical paths have been defined for the QCL beams. The first arm, delimited by a dashed line in Figure 1(a), is devoted to the active stabilization between the two QCLs by measuring the heterodyne beatnote on a commercial HgCdTe photodetector (Vigo Photonics) and then comparing it to a radiofrequency (RF) stable reference signal (Rohde&Schwarz SMC100A). A home-made PLL is used to actively control the current injected in the signal laser in order to phase-lock the latter to the LO. This results in a significantly narrowed beatnote signal down to the sub-Hz level [20]. The second arm of the setup in Figure 1(a) is used for the coherent detection of the signal by measuring the heterodyne beatnote with the LO on a QCD. QCDs [8,21] are unipolar infrared photodetectors that can operate in photovoltaic mode at room temperature. The active region is based on the periodic repetition of tunnel coupled quantum wells. Figure 1(b) represents one period of the QCD: mid-infrared photons are absorbed through an intraband transition between two confined states, labelled 1 and 2, of two tunnel coupled quantum wells [22]. The corresponding absorption spectrum is centred at the transition energy $E_{12} = E_2 - E_1$. Photoexcited electrons relax, by scattering with longitudinal optical phonons, towards the ground state of the following period of the cascade. The QCD operates in photovoltaic mode due to the asymmetry of the cascade region that acts as a pseudo-electric field driving the electrons in one direction only, thus giving rise to a photocurrent. Figure 1(c) shows the room temperature photocurrent spectra of the two QCDs used in our experiments. The detector used in the 4.8 μm set-up is a GaInAs/AlInAs QCD based on a diagonal design (Figure 1(b)), centred at 247 meV (5 μm), while the 9 μm detector, centred at 141 meV, is realized with a GaAs/AlGaAs heterostructure (the same device has been used in ref. [1]). The red dashed lines in Figure 1(c) indicate the emission energy of the QCLs used in our experiments, which are very close to the QCD photocurrent peaks. In order to operate at high frequency, the QCDs have been processed into 50 μm x 50 μm mesa structures that are electrically connected to a 50 Ω coplanar waveguide through an air-bridge for a low-inductance top contact. The device is then fixed on a custom-made sample-holder and wire-bonded onto a 50 Ω impedance matched coplanar waveguide. The room temperature characteristics of the QCDs are summarized in Table 1. More details on the device characterizations are provided in the supplementary material section.

## 3 Optical heterodyne detection with free running QCLs

Before studying the influence of the active stabilization on the coherent detection setup, we have characterized the heterodyne noise equivalent power with free-running QCLs. The stabilization arm indicated with dashed lines in Figure 1(a) is not used for this experiment. The heterodyne power is read on a radiofrequency spectrum analyser (Agilent E4407B) for a constant local oscillator optical power ($P_{LO}$ = 45 mW), while progressively decreasing the signal power, $P_S$, impinging on the detector by inserting optical density filters in the signal path. The heterodyne signal at ~ 130 MHz is amplified with a low noise transimpedance amplifier with a gain $G_{trans} = 5 \times 10^3$ V/A. Figure 2 presents the heterodyne power as a function of the signal power at 4.8 μm (panel (a)) and 9 μm (panel (b)), expressed in dBm. A linear behaviour is observed at both wavelengths over more than six orders of magnitude. In fact, starting from a signal power of 15 mW without optical densities, the minimum detectable signal power was 5 nW at 4.8 μm and 1.5 nW at 9 μm. The heterodyne current can be expressed as a function of the signal power as:

$$|I_{het}| = 2\mathcal{R}\eta\sqrt{P_{LO}P_S} \qquad (1)$$

where $\mathcal{R}$ is the QCD responsivity (given in Table 1) and $\eta$ is the heterodyne efficiency. The last parameter is an indicator of the quality of the optical setup, as it reflects the matching between signal and local oscillator in both amplitude and phase spatial distribution. The heterodyne power can be obtained from Eq. (1) by accounting for the transimpedance gain ($V_{het} = I_{het} \times G_{trans}$) and the instrument input load $R_L = 50\ \Omega$:

$$P_{het} = \frac{V_{het}^2}{R_L} = \frac{4\ \mathcal{R}^2 P_{LO}\eta^2 G_{trans}^2}{R_L} P_S \qquad (2)$$

The dashed lines in Figure 2 show the linear fit of the data (in logarithmic scale) by using Eq. (2). A slope of 0.997±0.008 is found for the experiment at 4.8 μm and 1.01±0.01 for the experiment at 9 μm. From the value of the intercept of the linear fit, we can extract the heterodyne efficiencies, which are found to be 55% at 4.8 μm and 36 % at 9 μm. In Figure 2, the measured noise level is indicated by a continuous line and the resolution bandwidth (RBW) is also provided. Identifying the origin of the noise is essential for a better understanding of the limiting factors of the heterodyne detection setup and possible improvements in the measurements. As they operate in photovoltaic mode, QCDs exhibit negligible dark current noise, which will not be considered in our estimations. Instead, we focus on thermal noise and generation – recombination noise [23, 24]. Thermal noise depends on the detector temperature, $T$, resistance $R_d$ and resolution bandwidth $\Delta f$:

$$i^2_{th} = \frac{4\ k_B\ T\ \Delta f}{R_d} \qquad (3)$$

The calculated values of thermal noise of the detector at room temperature are presented in Table 2. We also report the calculated values of the generation-recombination noise, obtained as

$$i^2_{ph} = 4egI_{ph}\Delta f$$

with $g$ representing the photoconductive gain, considered equal to the inverse number of periods in the absorbing region, and $I_{ph}$ denoting the photocurrent (for very low signal powers, $I_{ph} \simeq I_{LO}$). The values presented in Table 2 reveal that the sensitivity of our experiment is mainly set by the instrument noise floor. The NEP of the system, defined as the value of the power at which the signal to noise ratio is equal to 1, is obtained by linear extrapolation of the data till the noise level. We obtain 1nW at 4.8 μm and 200 pW at 9μm. In order to reduce the noise level and improve the sensitivity of the detection, we have used an active stabilization technique to reduce the resolution bandwidth to 1 Hz. As it will be seen in the following, the increase of the integration time induced by the stabilization also allows the measurement of the photocurrent without the need of an amplification stage.

## 4 Optical heterodyne detection with stabilised QCLs

The sensitivity of the heterodyne setup is limited by the stability of the beating, which sets the resolution bandwidth and consequently the integration time. Standard ways to reduce laser frequency fluctuations involve passive (i.e. minimizing temperature fluctuations, reducing the noise sources in the setup) or active stabilization of the QCLs. Several active stabilization experiments have been reported for linewidth narrowing, generally using a frequency discriminator, such as a high-finesse cavity [25] or a molecular resonance [26,27], to measure the laser frequency fluctuations and generate an error signal that is fed back into the applied current. Other stabilization techniques involve phase locking the quantum cascade laser to a stable frequency reference, such as a frequency comb [2,28], or compensating frequency fluctuations of the optical power in the QCL by using a laser diode [29].

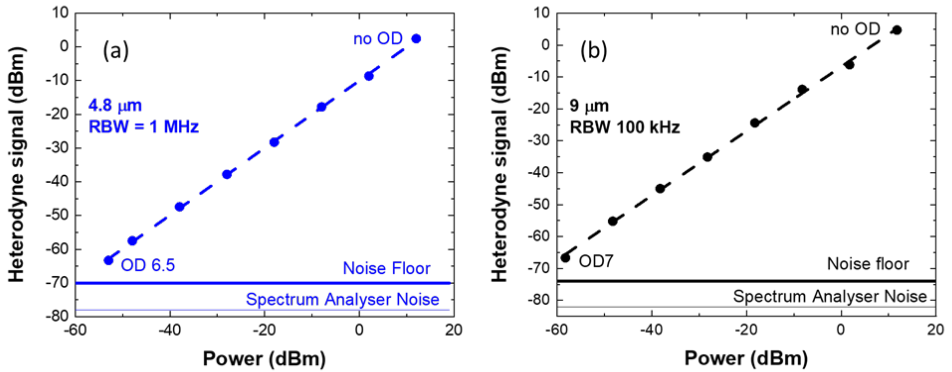

**Figure 2**. Measured heterodyne power at 4.8 μm (panel (a)) and 9 μm (panel (b)) as a function of the signal power after attenuation by optical densities. The resolution bandwidths (RBW) are indicated in the figures. The frequency of the heterodyne signal is 125MHz at 4.8 μm and 130 MHz at 9 μm. The QCDs are operated at room temperature in photovoltaic mode. The QCLs are free-running. The heterodyne current is amplified with a transimpedance amplifier. The dashed lines present linear fits to the data, allowing extraction of the heterodyne efficiencies, while the continuous lines indicate the noise floors. OD: optical density.

In our experiment, the beatnote between the signal and the local oscillator is stabilised by introducing a stabilization arm in the optical setup, as shown in Figure 1(a), which includes a second detector and a homemade PLL. The beatnote measured on the detector is sent to the PLL and compared to a reference RF signal, and the PLL is used to measure the phase difference between the two signals and actively control the injection current of the signal laser to stabilise the beatnote. It is important to underline that for this experiment no amplification stage has been used, and that all the devices are operated at room temperature. Figure 3 presents the heterodyne signals measured with a RBW of 1 Hz with the two stabilized setups at 4.8 μm (panels (a) and (b)) and 9 μm (panels (c) and (d)). In 3(a) and 3(c), the different spectra are obtained by progressively attenuating the signal optical power. It is important to underline that the full width at half maximum (FWHM) extracted through a Gaussian fit is very close to the RBW = 1 Hz (figure 3(b) and 3(d)).

The attenuation of the heterodyne peak power for different signal powers is shown in Figure 4. The detected signal now displays a linear dynamic range over more than 9 orders of magnitude, with a slope 0.974±0.008 at 4.8 μm and 1.005±0.008 at 9 μm. When no amplification is present, the heterodyne voltage drops on the load resistance of the instrument

$$P_{het} = R_L \times I_{het}^2 = 4\, R^2\, P_{LO}\, \eta^2\, R_L\, P_S \qquad (4)$$

and the values of the heterodyne efficiency extracted from the intercept of the linear fit are 50% at 4.8 μm and 80% at 9μm Note that in the 9 μm set-up the stabilization technique has allowed a more careful alignment of the signal and LO on the detector, resulting in an improved heterodyne efficiency. The lowest power at which the heterodyne signal could be measured is 2 pW (resp. 15 pW) at 4.8 μm (resp. 9 μm), corresponding to $5 \times 10^7$ photons per second (resp. $7 \times 10^8$ photons per second). Table 3 reports the different noise contributions for a RBW of 1 Hz. The thermal noise has been measured by connecting the detectors to a spectrum analyser (Zurich Instrument UHFLI) after proper amplification to surpass the instrument noise floor, in dark operation. The measured values are systematically higher than those estimated by using Eq. (3). This discrepancy could be attributed to the use of an amplification stage for the measurement. Table 3 shows that in this experiment the spectrum analyser constitutes again the main source of noise. However, we should note that, in the 9 μm experiment, our measurements were affected by pick-up noise (black dotted line in Figure 4), probably due to the packaging of the laser.

Despite the presence of the pick-up noise at 9 μm, the lowest measured signal power of 15 pW is lower than the NEP extrapolated at 1Hz in ref. [30].

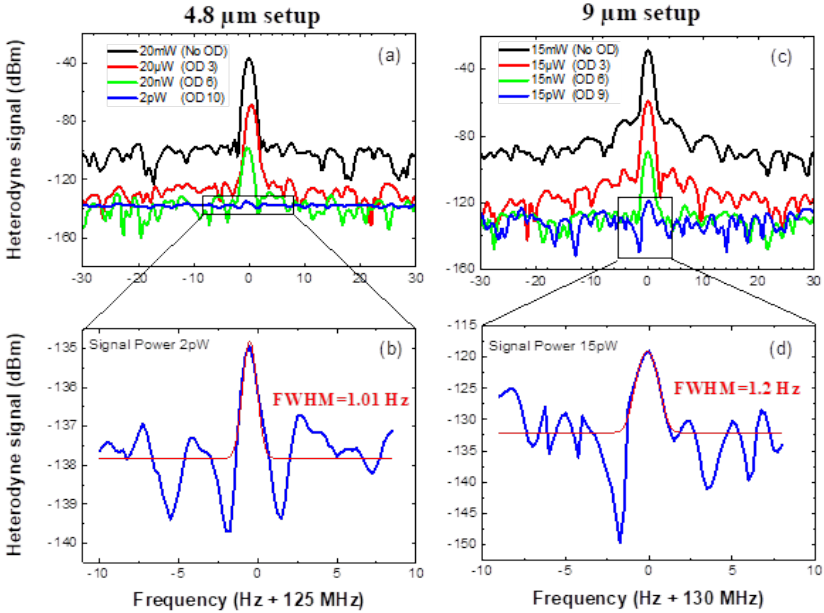

**Figure 3**. Top panels: Beat-note measured for different attenuated signal powers as indicated in the legend at a RBW of 1 Hz (left and right: 4.8 μm and 9 μm setup, respectively). Lower panels: Gaussian fit (red line) of the heterodyne signal (RBW=1 Hz). OD: optical density. FWHM: full width at half the maximum.

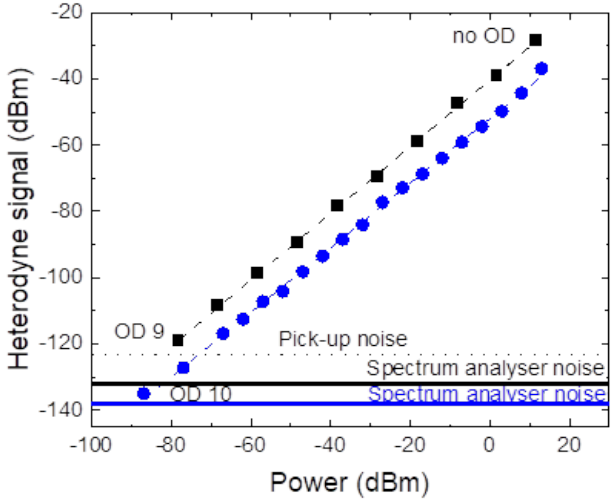

**Figure 4.** Stabilized setups. Heterodyne power measured at 4.8 µm (blue dots) and 9µm (black squares) as a function of the optical power of the signal after attenuation. The RBW is 1 Hz and the heterodyne frequency is 125 MHz at 4.8 µm and 130 MHz at 9 µm. Dash-dotted lines are linear fits following Eq. (4). Solid lines indicate the instrument noise floors, while the dotted black line shows the level of electrical pick-up noise in the 9 µm experiment.

In the absence of pick-up noise, at 9 µm a NEP of 760 fW (corresponding to $3.4 \times 10^7$ photons per second) at 9µm is expected by intercepting the data curve and the noise floor set by the spectrum analyser. Note that the introduction of an amplifier in the setup did not improve the signal-to-noise ratio of the heterodyne detection, and therefore it was not included in the experiment. Table 4 contains a summary of the values of the NEP obtained with the different techniques, including direct detection. We can see that, remarkably, the heterodyne technique allows a strong improvement of the sensitivity of the detection in both atmospheric windows, as the NEP is reduced by more than 6 orders of magnitude with respect to direct detection.

As the heterodyne frequency used in this experiment is lower than the high frequency cut-off of a typical Peltier cooled (operating at 200 K) commercial Mercury Cadmium Telluride (MCT) detector, one could ask how our room temperature QCD and MCTs compare as heterodyne receivers (see Supplementary Information). We estimate that, considering the same heterodyne efficiency as in our experiment, the NEP expected from a heterodyne set-up employing a commercial MCT detector could be 40 dBm lower than the one we measured. This difference could be easily compensated by pre-amplifying the QCD, as the noise level is not limited by the device (see Table 3). When using the QCD as heterodyne receiver, the NEP could be further reduced if a waveguide [31] or metamaterial [32] architecture is employed for the device. Furthermore, the local oscillator power can be increased up to a few hundreds of mW when using the QCD, while MCTs usually saturate for an impinging power of few hundreds µW. QCDs have, therefore, the full potential to be high sensitivity heterodyne receivers operating room temperature and with frequency response of few tens of GHz [1].

## 5 Conclusion

In conclusion, we exploited the peculiar characteristics of unipolar quantum devices combined with an active stabilization technique, to demonstrate heterodyne detection in the mid-infrared atmospheric windows with pW sensitivity. The detectors operate at room temperature, without any amplification stage and show a linear behaviour for more than 9 orders of magnitude.

In order to estimate the maximum achievable sensitivity of the heterodyne setup, we consider the limit where the most important contribution to the noise comes from the generation - recombination noise, and where the signal power is negligible with respect to that of the local oscillator. In this limit, and considering unity heterodyne efficiency, the signal-to-noise ratio can be written as:

$$\frac{S}{N} = \frac{2\mathcal{R}\sqrt{P_{LO}P_S}}{\sqrt{4\, eg\, \Delta f\, \mathcal{R}\, P_{LO}}} = \sqrt{\frac{\alpha P_s}{E\, \Delta f}}$$

Where the responsivity has been expressed in terms of the absorption quantum efficiency, $\alpha$, and the photon energy, $E$, as: $\mathcal{R} = e\alpha g/E$. The minimum achievable NEP is thus:

$$NEP = \frac{E\Delta f}{\alpha} \qquad (5)$$

At 9 μm wavelength, the NEP associated with the detection of one photon per second corresponds to 0.02 aW. Equation (5) shows that the heterodyne signal is ultimately limited by the absorption quantum efficiency of the detector, and by the resolution bandwidth. In this work, we have been focusing on the reduction of the bandwidth through the active stabilisation of the lasers. An interesting perspective to improve the absorption quantum efficiency concerns the use of photodetectors based on patch antenna resonators [15, 30]. They have already been successfully implemented in heterodyne detection setups and can be packaged in high speed architectures [17] and designed to operate in the critical coupling regime [31]. Furthermore, they operate in reflectivity rather than in transmission, which could be an advantage for optical alignment.

## 6 Authors' statements

**Research funding:** The authors acknowledge financial support by the ENS-Thales Chair, by PEPR Electronique, by the Agence Nationale de la Recherche (project CORALI ANR-20-CE04-0006 and project COLECTOR ANR-19-CE30-0032) and by H2020 Future and Emerging Technologies (Project cFLOW).

**Authors contribution:** M.S. and L.D.B. performed all the experiments under the supervision of D.G, A.V and C.S.. M.S, L.D.B., D.G., A.V., and C.S. performed the data analysis and wrote the manuscript. Y.T. contributed to valuable discussions. E.R. fabricated the QCDs. O.L. and B.D. fabricated the PLL. L.L., A.G.D. and E.L. performed the samples growth. A.V. and C.S. directed the research.

**Conflict of interest:** Authors state no conflict of interest.

Table 1 Main characteristics of the quantum cascade detectors used in the experiment.

| | Material system | Number of periods | Photocurrent peak (meV) | Responsivity $\mathcal{R}$ (mA/W) | 3-dB cut-off frequency (GHz) | $R_d$ (Ω) |
|---|---|---|---|---|---|---|
| **4.8µm QCD** | GaInAs/AlInAs | 12 | 247 | 1.9 | 2 | 1.8 kΩ |
| **9µm QCD** | GaAs/Al$_{0.35}$Ga$_{0.65}$As | 12 | 141 | 4.2 | 3 | 180 Ω |

Table 2. Different noise contributions for 4.8 μm and 9 μm QCDs in the case of free running QCLs. The calculation includes the transimpedance amplifier gain.

|  | RBW (MHz) | Calculated thermal noise (dBm) | Calculated generation-recombination noise (dBm) | Spectrum analyser noise (dBm) |
|---|---|---|---|---|
| **4.8 μm QCD** | 1 | - 83.5 | - 86.4 | - 78 |
| **9 μm QCD** | 0.1 | - 83.5 | - 93 | - 82 |

**Table 3.** Different noise contributions for 5 μm and 9 μm QCDs at RBW = 1Hz.

|  | Thermal noise | Calculated generation-recombination noise | Spectrum analyser noise |
|---|---|---|---|
| **4.8 μm QCD** | -183.5 dBm (calculated)<br>-176 dBm (measured) | -186.4 dBm | - 138 dBm |
| **9 μm QCD** | -173.5 dBm (calculated)<br>-168 dBm (measured) | - 183 dBm | - 132 dBm |

**Table 4**. Summary of the values of the Noise Equivalent Power (NEP) obtained with direct detection and heterodyne detection, with free running and stabilized QCLs for the two wavelengths.

| | Direct detection | Free running QCLs and amplification | Stabilised beatnote (no amplification) |
|---|---|---|---|
| **NEP @ 4.8 µm (W)** | 3.7 x 10$^{-6}$ | 9.6 x 10$^{-10}$ | 1.5 x 10$^{-12}$ |
| **NEP @ 9µm (W)** | 1.3 x 10$^{-6}$ | 2 x 10$^{-10}$ | 6 x 10$^{-12}$ (7.6 x 10$^{-13}$ without pick-up noise) |

# Ultra-sensitive heterodyne detection at room temperature in the atmospheric windows

## Supporting information

Mohammadreza Saemian[1,*], Livia Del Balzo[1,*], Djamal Gacemi[1], Yanko Todorov[1], Etienne Rodriguez[1], Olivier Lopez[2], Benoit Darquié[2], Lianhe Li[3], Alexander Giles Davies[3], Edmund Linfield[3], Angela Vasanelli[1,#], Carlo Sirtori[1]

[1]Laboratoire de Physique de l'Ecole Normale Supérieure, ENS, Université PSL, CNRS, Sorbonne Université, Université Paris Cité, 75005 Paris, France

[2]Laboratoire de Physique des Lasers, CNRS, Université Sorbonne Paris Nord, 93430 Villetaneuse, France

[3]School of Electronics and Electrical Engineering, University of Leeds, Woodhouse Lane, Leeds LS2 9JT, UK

*These authors contributed equally to this work.
#Corresponding author: angela.vasanelli@ens.fr

## 1. Description of the quantum cascade detectors

The quantum cascade detector (QCD) used in the 4.8 µm set-up, centered at 5 µm, is composed of 12 periods of lattice matched InGaAs wells and InAlAs barriers, grown on an InP substrate by Molecular Beam Epitaxy. In a single period, the thicknesses, in nm, of the InGaAs wells and AlInAs barriers (indicated in bold) are the following: <u>3.8</u>/**1.8**/0.9/**6.2**/1.3/**5.5**/1.7/**5.0**/2.4/**3.8**, with the first underlined layer being an InGaAs quantum well, doped at $3 \times 10^{18}$ cm$^{-3}$. The conduction band diagram with the square moduli of the relevant wavefunctions, plotted at the corresponding energies, is shown in Figure 1(b) in the main text.

The 9 µm QCD structure is composed of 12 periods of GaAs wells and Al$_{0.35}$Ga$_{0.65}$As barriers grown by Molecular Beam epitaxy on a GaAs substrate. The single period structure is (all the thicknesses are in nm): <u>4.4</u>/**1.4**/1.4/**5.5**/1.7/**5.8**/2.3/**5.2**/3.0/**4.8** with the first underlined layer being GaAs well doped at $1 \times 10^{18}$ cm$^{-3}$, while the barrier thicknesses are indicated in bold. The conduction band diagram with the square moduli of the relevant wavefunctions, plotted at the corresponding energies, is shown in Figure S1.

The QCDs were processed into high-frequency coupled mesas[1,2,3]. Square mesa regions of 50 µm size were defined through optical lithography. In the case of the 5 µm QCD, the mesas were chemically etched down to the bottom contact layer using H$_3$PO$_4$:H$_2$O$_2$:H$_2$O (1:1:38). Then the half plane starting just below the mesa was protected with resist, allowing the contact to be etched away (using the same H$_3$PO$_4$ solution) down to the substrate. After depositing a sacrificial support of reflowed S1818 resist for the air bridge, a Ti/Au 50 Ω coplanar waveguide was evaporated in one step using negative resist AZ5214 E for patterning. A photoresist stripper (SVC 14) was preferred to the standard acetone lift off to ensure that the bridge was properly freed from resist.

In the case of the 9 µm QCD, physical etching instead of wet chemical etching, was used, as explained in ref. [1].

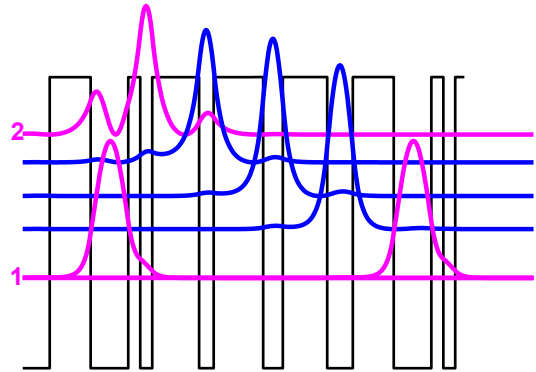

*Figure S1 : Band structure of the 9 µm QCD. The square moduli of the wavefunctions involved in the photon absorption are indicated in purple, while those involved in the electron extraction are indicated in blue.*

## 2. QCD responsivity

To measure the responsivity of the devices, a DFB QCL ($\lambda$=4.8 µm from Ad Tech Optics and $\lambda$=9 µm from Thorlabs) is focused on the active area of the QCD with a F=12.5 mm lens. The generated photocurrent is read on a sourcemeter (Keithley 2450), while the incident power is measured with a powermeter by means of a flipping mirror that can deviate the optical path (experimental setup in Figure S2).

Figure S3 shows the photocurrent of the two devices measured at room temperature as a function of the QCL power impinging on the detector. The responsivity extracted from the slope is 1.9 mA/W for the 5 µm device and 4.2 mA/W for the 9 µm device.

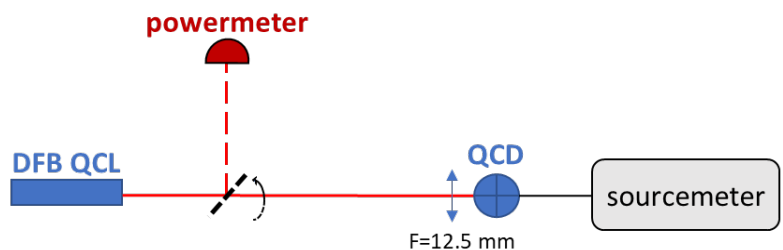

*Figure S2. Sketch of the setup for responsivity measurements.*

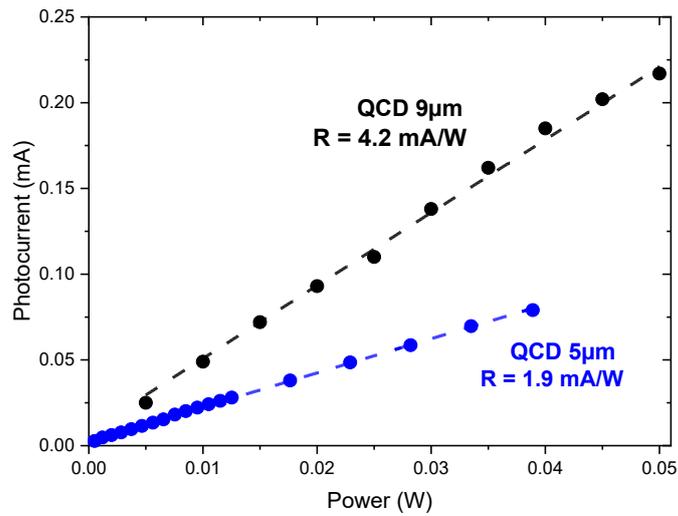

*Figure S3. QCDs photocurrent vs laser power. Dashed lines show the linear fit of the data, from which the responsivity is extracted.*

## 3. QCDs frequency response

The frequency response of the QCDs was measured by using the rectification technique, that relies on the non-linear current-voltage (I-V) characteristic of the detector [4]. The experimental setup is shown in Figure S4. A sinusoidal radiofrequency signal generated by a synthesizer (Anritsu MG3693B) is sent to the QCD through the AC port of a bias-tee. The rectified current, extracted through the DC port of the bias-tee, is measured on a sourcemeter (Keithley 2450). The rectified current, which is proportional to the squared voltage transfer function, $H(\omega)$ of the device, is collected while sweeping the frequency of the AC signal up to 30 GHz. Figure S5 shows the rectified power (i.e. the squared rectified current times the load resistance) as a function of the AC frequency for the 5 µm detector. A 3dB cutoff frequency of 2 GHz is extracted from this plot. Similar measurements were performed for the 9 µm QCD, and the corresponding results have been presented in ref. [1]. The measured cutoff frequency for the 9 µm detector was 5 GHz.

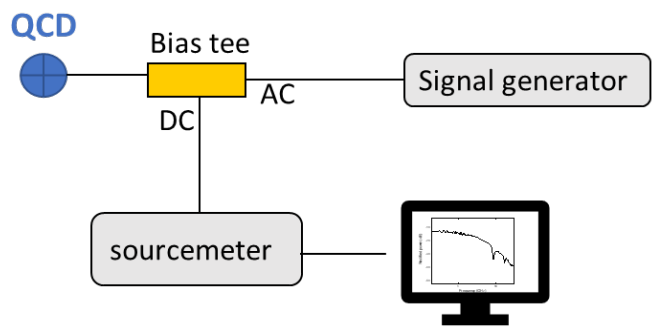

*Figure S4: Experimental setup for rectification measurements.*

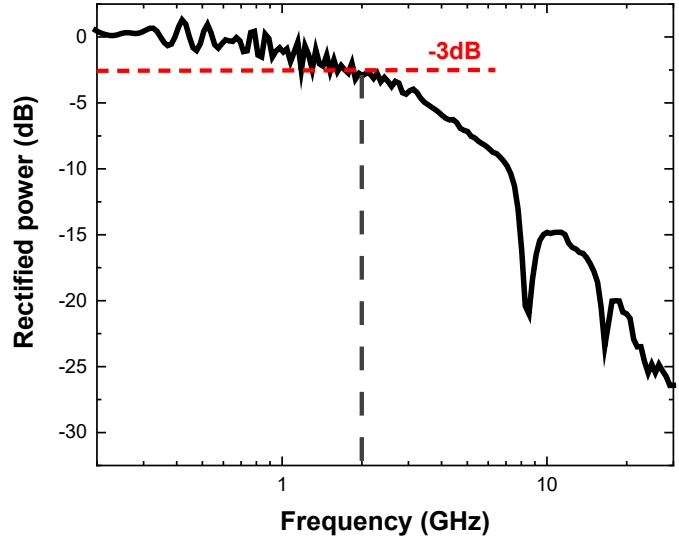

*Figure S5. Rectified power plotted as a function of the AC frequency for the 5 µm QCD.*

## 4. Estimated NEP of the QCDs in direct detection

The Noise Equivalent Power (NEP) of the QCDs in direct detection is:[5]

$$NEP = \frac{i_n}{\mathcal{R}}$$

where $i_n$ is the current noise and $\mathcal{R}$ responsivity of the detector (see section 2). In direct detection, the measurement bandwidth is limited by the QCL linewidth, which is 1 MHz at 4.8 µm and 100 kHz at 9 µm. The most important contribution to the noise is the thermal noise, which is estimated to be -116 dBm for the 4.8 µm QCD and -118 dBm for the 9 µm QCD. This estimation is performed starting from the measured thermal noise power at 1Hz (see main text), and considering the measurement bandwidth. From these values it is possible to deduce the current noise as:

$$i_n = \sqrt{\frac{P_{thermal\ noise}\ (Watt)}{50\ \Omega}}$$

We finally obtain a NEP of 3.7 µW at 4.8 µm and 1.34 µW 9 µm.

|  | Measured thermal noise power at 1Hz | Thermal noise intensity in direct detection | Thermal current noise $i_n$ | NEP in direct detection |
|---|---|---|---|---|
| 4.8 µm QCD | -176 dBm | -116 dBm @1MHz | 7.1 nA | 3.7 µW |
| 9 µm QCD | -168 dBm | -118 dBm @100kHz | 5.6 nA | 1.34 µW |

*Table S1. Estimation of the NEP of the QCD detectors. The thermal noise is measured with a measurement bandwidth of 1Hz (see main text). The thermal noise intensity is then estimated by considering the measurement bandwidth set by the linewidth of the QCLs. The thermal current noise and then the NEP are finally extracted from this value.*

# 5. Comparison with a commercial Peltier cooled Mercury Cadmium Telluride detector

In this section we compare the capabilities of our QCD as heterodyne receiver at 9 µm with respect to a commercial Peltier cooled Mercury Cadmium Telluride (MCT) detector (Vigo PVI-4TE-10.6-0.5x0.5).

The characteristics of the two devices that are relevant for heterodyne detection are summarized in Table S2.

|  | Operation temperature | High cut-off frequency | Responsivity | Pre-amplification | Noise level @ 1Hz | Saturation power |
|---|---|---|---|---|---|---|
| Peltier cooled MCT | 200 K | 700 MHz | 2.55E+4 V/W | Yes (8.65E+3 V/A) | -120 dBm | 200 µW |
| QCD | 300 K | 3 GHz | 4 mA/W | Non | -168 dBm | > 100 mW |

*Table S2. Comparison of the measured characteristics of a commercial Peltier cooled MCT detector and of our 9 µm QCD.*

The QCD is a photovoltaic device that does not need to be cooled. As it operates at 0 V and it is not amplified, its noise level is typically lower than the MCT's one. At room temperature the QCD responsivity is of few mA/W, three orders of magnitude lower than the MCT's one. One should note that, contrarily to the QCD, the MCT is pre-amplified, with a transimpedance gain of $10^4$ V/A. The responsivity of QCD can be increased by using metamaterial - based devices (50 mA/W at 0V and room temperature in ref. [6]) or waveguide geometries (400 mA/W in ref. [7]). The QCD can also be pre-amplified, as we did in the experiment presented in section 3 of the main manuscript.

An important property for heterodyne detection is the saturation power of the detector. For the MCT detector, the saturation power is 200 µW. In the case of the QCD, we measured a linear behaviour of the photocurrent as a function of the incident QCL power up to 100 mW (the maximum power delivered by the QCL)-. However, the theoretical saturation intensity of the QCD could be much higher, as in an intersubband system at ~10 µm it can reach 1MW/cm².[8] This means that a local oscillator with a power up to three orders of magnitude higher than that used in our experiment could be employed if an intersubband detector is used as local oscillator.

Finally, an important difference between the two devices is their frequency bandwidth. Figure S6 presents a direct optical measurement of the frequency bandwidth obtained by shining a mid-infrared frequency comb (Menlo System FC1500-ULN) onto the detector. The beating between the optical teeth appears as beatnotes separated by 100 MHz. From this figure, we extract a -3dB cut-off of 3 GHz for the 9 µm QCD, and of 700 MHz for the MCT detector.

In the experiment discussed in the main text, the heterodyne signal is at 130 MHz, limited by the PLL cut-off frequency. As this frequency is lower than the high frequency cut-off of the MCT detector, one could ask whether a commercial Peltier cooled MCT could provide a lower NEP than the QCD in a heterodyne detection experiment. We estimate that, assuming the same heterodyne efficiency as in our experiment, the NEP expected from a heterodyne set-up employing the commercial detector presented above could be 40 dBm lower than the one we measured. This difference could be easily compensated by pre-amplifying the detector, as the noise level is not limited by the device.

Several improvements can be introduced in the QCD architecture and packaging in order to reduce the NEP. Indeed, one can use a QCD based on a waveguide geometry, which can be also pre-amplified and cooled by using a packaging similar to the MCT's one. Furthermore, the local oscillator power can be increased up to a few hundreds of mW.

Considering a QCD detector with 0.5A/W responsivity (which is feasible if considering a Peltier cooled device in a waveguide geometry) and a local oscillator power of 200 mW, we estimate a noise equivalent power of 10 aW, corresponding to the detection of 500 photons per second.

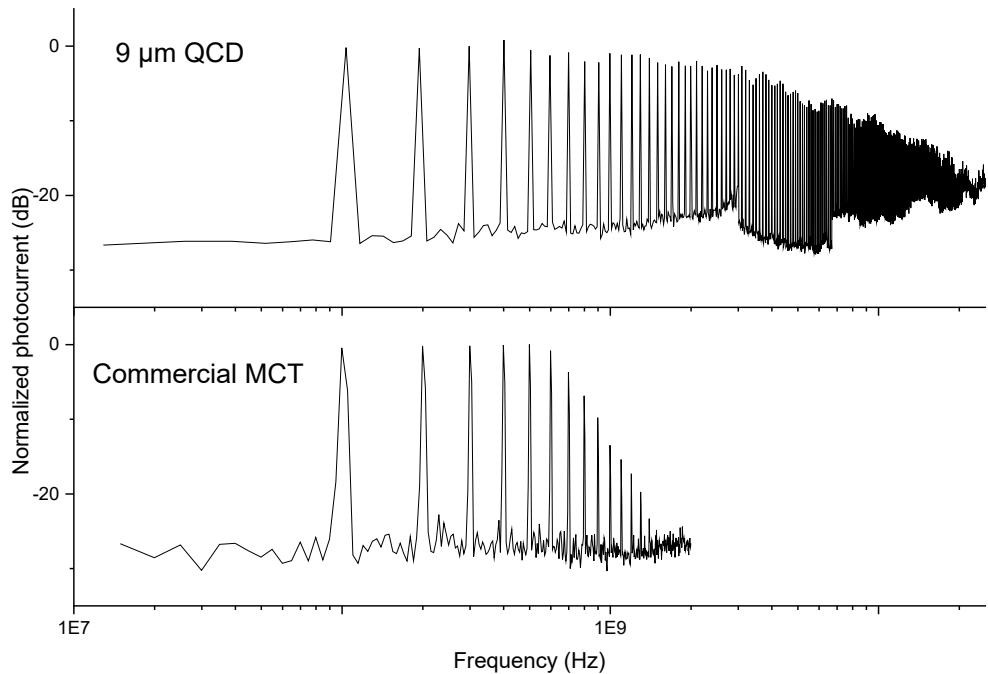

*Figure S6. Normalized photocurrent measured on the 9 μm QCD and on the commercial MCT by using a mid-infrared frequency comb with teeth separated of 100 MHz.*